\documentstyle[preprint,aps,epsfig]{revtex}
\begin{document}
\draft 
\title{Matter effects for `just so' oscillations}
\author
{Mohan Narayan and S. Uma Sankar}
\address
{Department of Physics, I.I.T. , Powai, Mumbai 400076, India}
\date{\today}
\maketitle

\begin{abstract}
We study the effect of the matter term on the evolution of the 
solar neutrinos when the neutrino parameters are those of the 
`just-so' case. The extreme non-adiabatic effects at the edge 
of the sun reduce the expression for the survival probability 
in the just-so case to that of the vacuum case. This conclusion
is independent of the width of the extreme non-adiabatic region,
which is a function of the density profile of the sun beyond
$r > 0.9 R_s$. However, in its propagation, neutrino encounters
regions of moderate (non-extreme) non-adiabticity. Neutrino 
traversal through these regions give corrections to the survival
probability which are profile dependent. 

\end{abstract}
\vspace{0.5cm}
\narrowtext

\section{Introduction}
The solar neutrino data of Super-Kamiokande has lead to a reexamination 
of all the solutions of the solar neutrino problem \cite{sksol}. In 
addition to the overall suppression rate, Super-Kamiokande has measured
the day-night asymmetry and the spectrum of the scattered electrons.
These two measurements are independent of the overall normalization of
the $^8$B neutrino flux. This normalization has the largest uncertainty
among all the predictions of the solar models \cite{bahcall,bp95}. 
Hence data, which are independent of it, are extremely important in 
studying the properties of $^8$B neutrinos from the sun. The present data,
especially the recoil spectrum, 
favor `just-so' oscillation solution to the solar neutrino problem 
\cite{sksol,bks98,smy}.

In analyzing the solar neutrino data in terms of `just-so' oscillations,
the expression used for electron neutrino survival probability $P_{ee}$ 
is simply the vacuum survival 
probability. Given that the matter term in the sun is several orders
of magnitude greater than the mass-squared difference, the question
arises whether the use of vacuum survival probability is justified.
A very simple justification can be given the following way: 
The mass-squared difference $\delta$ required for just-so oscillations
is about $10^{-10}$ eV$^2$. The matter term 
\begin{equation}
A~({\rm in~eV}^2) = 0.76 \times 10^{-7} 
\rho~({\rm in~gm/cc}) E~({\rm in~MeV}),
\end{equation}
is much larger than $\delta$ as long as the density $\rho$ is greater
than $0.01$ gm/cc. First we assume that the density is spherically symmetric
and it falls abruptly to $0$ once it decreases to a value of $0.01$ gm/cc.
As long as the neutrino is in the sun, $A$ completely overwhelms $\delta$
and the electron neutrino in the sun is
essentially the higher mass eigenstate. Hence during its travel 
through the sun, the neutrino does not oscillate and it simply acquires
a phase. Thus throughout the travel through the sun, an electron neutrino 
remains an electron neutrino. When the neutrino comes out of the sun, it
passes through the abrupt change in density from $0.01$ to $0$. In such a 
case, the flavour composition is unchanged. Since the neutrino in the sun 
remained an electron neutrino, we have an electron neutrino coming
out of the sun. Thus the starting point of neutrino evolution is
transferred from the core of the sun to the edge of the sun \cite{thpaul}. 

But now we can raise the question: Suppose the solar matter density
does not change abruptly to zero but goes to zero smoothly. Then at
some point in its propagation the neutrino will pass through a 
region where the density is equal to the resonant density. In such 
a situation, how is the neutrino oscillation probability modified?
We address the question below.

\section{Just-so Survival Probability}
For simplicity here we consider only two flavor oscillations. The flavor 
states $\nu_e$ and $\nu_\mu$ are linear combinations of the two mass
eigenstates $\nu_1$ and $\nu_2$
\begin{eqnarray}
\nu_e & = & \cos \theta~\nu_1 + \sin \theta~\nu_2 \nonumber \\
\nu_\mu & = & - \sin \theta~\nu_1 + \cos \theta~\nu_2.
\end{eqnarray}
Without loss of generality we can take $\nu_2$ to be more massive than
$\nu_1$ and hence $\delta = m_2^2 - m_1^2 >0$. If $\theta < \pi/4$, then
$\nu_e$ is predominantly the lighter state. If $\theta >\pi/4$, then 
$\nu_e$ is predominantly the heavier state. Thus the two physically 
distinguishable possibilities are covered by taking the range of $\theta$
to be $(0,\pi/2)$. Since we are considering `just-so' oscillations, we
will assume that $\delta = 10^{-10}$ eV$^2$. We will also assume that 
the value of $\theta$ is moderately large, {\it i.e.} $\theta$ is not
finetuned to be very small or be very close to $\pi/4$.  

Electron neutrinos are produced in the core of the sun. 
Since the solar matter near the core is very dense, one must take the
matter term $A$ into account in determining the mass eigenstates in 
the core.
We can define instantaneous matter dependent mass eigenstates
\begin{eqnarray}
\nu_e & = & \cos \theta_m~\nu_{1m} + \sin \theta_m~\nu_{2m} \nonumber \\
\nu_\mu & = & - \sin \theta_m~\nu_{1m} + \cos \theta_m~\nu_{2m},
\end{eqnarray}
where $\theta_m$ is the matter dependent mixing angle and is given by
\begin{equation}
\cos 2 \theta_m = \frac{\delta \cos 2 \theta - A}{\delta_m}.
\end{equation}
$\delta_m$ is the matter dependent mass-square difference and is given by
\begin{equation}
\delta_m = \sqrt{(\delta \cos 2 \theta - A )^2 +
(\delta \sin 2 \theta)^2}.
\end{equation}
As the neutrino travels through the sun, it encounters matter of 
constantly decreasing density. In the basis of matter dependent 
mass eigenstates, the evolution equation for the neutrino is 
\begin{equation}
i \frac{d}{dt} \left[ \begin{array}{c} \nu_{1m} \\ \nu_{2m} 
\end{array} \right]
= \frac{1}{4E} \left[ \begin{array}{cc} 
- \delta_m (t) & -4 i E \dot{\theta}_m (t) \\
4 i E \dot{\theta}_m (t) & \delta_m (t) \end{array} \right]
\left[ \begin{array}{c} \nu_{1m} \\ \nu_{2m} 
\end{array} \right].
\label{eq:eqprop}
\end{equation}
The propagation of neutrino is adiabatic as long as the off-diagonal
terms are smaller than the difference of the diagonal terms in the above 
equation, {\it i.e.} 
\begin{equation}
\delta_m \gg \frac{2 E A \delta \sin 2 \theta}{\delta_m^2}
	     \left| \frac{\dot{A}}{A} \right|.
\label{eq:adbcon1}
\end{equation}
In the body of the sun the density profile is of the form 
\begin{equation}
\rho (r) = \rho_0 \exp (-10.54 r/R_s) \label{eq:expden}
\end{equation}
where $\rho_0$ is the core density (about 140 gm/cc) and $R_s$ is the 
radius of the sun (about $7 \times 10^5$ km) \cite{bahcall}.
Therefore
\begin{equation}
\left| \frac{\dot{A}}{A} \right| = 
\left| \frac{\dot{\rho}}{\rho} \right| = \frac{10.54}{R_s}  
= 3 \times 10^{-15}~{\rm eV}.
\end{equation}
As long as $\rho \geq 0.01$ gm/cc (or $r \leq 0.9 R_s$) the 
matter term $A$ is much larger than $\delta$, hence $\delta_m \simeq A$.
According to our assumption on $\theta$, $\sin 2 \theta \sim 1$.  
Substituting all these in Eq.~(\ref{eq:adbcon1}), 
the adiabatic condition becomes
\begin{equation}
A^2 \gg 6 \times 10^{-19} (E/{\rm MeV}).
\end{equation}
The most compelling reason for considering just-so oscillations as a 
solution to the solar neutrino problem is Super-Kamiokande data on 
electron spectrum due to $^8$B neutrinos. The observed range of these
neutrinos is 6-14 MeV and they are peaked around 10 MeV. Henceforth,
we will take E = 10 MeV for illustrative purpose. 
Conservatively we require that $A^2$ should be at least 10 times
the value on RHS. With this requirement, we find that the adiabatic
condition is satisfied if 
\begin{equation}
\rho \geq 0.01~{\rm gm/cc}. \label{eq:rhobnd}
\end{equation}
The adiabatic condition is likely to break down if $\rho \leq 
0.01$ gm/cc, or for radial distances greater than $r > 0.9 R_s$.

When the neutrino is well out of the sun the matter term is much  
smaller than $\delta$ and adiabaticity is restored.
In such a case we have, essentially, vacuum propagation. The radial
distance at which this happens is a function of the density profile
of the sun beyond $0.9 R_s$. For example, if the density falls
linearly beyond $0.9 R_s$, vacuum propagation starts around $r = R_s$.
If the density fall continues to be the exponential form shown in
Eq.~(\ref{eq:expden}), then vacuum propagation starts only around $2 R_s$.
Between the time of breakdown of the adiabaticity and the time of its
restoration, the off-diagonal terms in the equation of propagation,
Eq.~(\ref{eq:eqprop}), are comparable to the diagonal terms. In fact,
for some intermediate region, the off-diagonal terms completely dominate,
making the propagation extremely non-adiabatic. These extreme non-adiabatic
effects cause the just-so oscillation probability to reduce to 
vacuum oscillation probability.

An electron neutrino is produced in the core of the sun at time $t_0$
and proipagates adiabatically upto $t_1$. Between times $t_1$ and $t_2$
its propagation is non-adiabatic. Beyond $t_2$ the neutrino propagates
in vacuum and is detected at $t_3$. Its state vector at $t_3$ is  
\begin{equation}
| \Psi_e (t_3) \rangle =
\sum_{j,i}  |\nu_j \rangle exp \left(-i \varepsilon_{j} (t_3 - t_2)
\right) M_{ij} 
exp \left(-i \int_{t_0}^{t_1} \varepsilon^S_{i}(t) dt \right) 
U^C_{e i},
\label{psiatt3}
\end{equation}
where $\varepsilon^S_{i} (t)$ are the matter dependent energy eigenvalues
in the sun and  $\varepsilon_{i}$ are the energy eigenvalues in vacuum. 
$M_{i j}$ is the amplitude for non-adiabatic transition from
mass eigenstate $\nu_{im}$ at $t=t_1$ to mass eigenstate $\nu_j$ at $t=t_2$. 
$U^C_{ei}$ is the $e-i$ element of the matter dependent mixing matrix 
at the core of the sun. From the above equation, we obtain the electron
neutrino survival probability to be 
\begin{eqnarray}
P_{e e} &=& |\langle \nu_e |\Psi_e (t_3) \rangle|^2 
\nonumber \\ 
&=& \sum_{j', i'} \sum_{j i}
U_{e j} M_{i j} U^C_{e i} 
U_{e j'} M_{i' j'} U^C_{e i'} 
exp \left\{i\left(
\int_{t_0}^{t_1} (\varepsilon_{i'}^{S}(t) - \varepsilon_{i}^{S}(t)) dt 
\right)\right\}  
exp \left\{i(\varepsilon_{j'} -\varepsilon_{j})
(t_{3}-t_{2})\right\}. 
\end{eqnarray}
Because of the variation in the production region, the phase picked up 
in the time from $t_0$ to $t_1$ can be averaged out. So $exp\left(i
\int_{t_0}^{t_1} (\varepsilon_{i'}^{S}(t) - \varepsilon_{i}^{S}(t)) dt
\right)$ 
can be replaced by $\delta_{i i'}$. Thus, $P_{ee}$ is simplified to
\begin{equation}
P_{e e} = 
\sum_{i j j'}
U_{e j} M_{i j} U^C_{e i} 
U_{e j'} M_{i j'} U^C_{e i} 
 exp \left\{i(\varepsilon_{j'} -\varepsilon_{j}) (t_3 - t_2)
\right\}. 
\end{equation}
The mixing angle at the core of the sun $\theta_C = \pi/2$, because 
the electron neutrino is the heavier mass eigenstate due to the 
dominance of the matter term. In vacuum, of course, it is $\theta$.
Substituting these in the expression for $P_{ee}$, we get
\begin{eqnarray}
P_{ee} &=& x_{12} \cos^2 \theta  + (1 - x_{12}) \sin^2 \theta  \nonumber \\
&+& 2 \sin \theta \cos \theta 
\Re \left(M_{12}^*M_{22} \right) 
\cos \left( 2.54 \frac{\delta~x}{E} \right), \label{eq:pee3}
\end{eqnarray}
where $x_{12} = |M_{12}|^2$ is the probability for the $\nu_{1m}(t_1)$ 
evolve into $\nu_2$ at $t_2$. In the above equation we have also made
the extreme non-relativistic apporoximation for neutrinos and replaced
the difference in energies by the mass-squared difference and the time
time of travel $(t_3 - t_2)$ by the distance of travel $x$. 

The survival probability given in Eq.~(\ref{eq:pee3}) should give us
vacuum survival probability if the density falls abruptly to $0$,
that is $t_2 - t_1$ is extremely small. In such a case, it was shown
that $M_{12} = (U_S^\dagger U)_{12} = \sin (\theta_S - \theta) =
\cos \theta$ because $\theta_S$, the mixing angle in the sun, is $\pi/2$
\cite{kuopan}.
For $M_{12} = \cos \theta$, $P_{ee}$ immediately reduces to vacuum 
survival probability. Now the question is: what is $M_{12}$ if the 
density falls smoothly to zero and the distance between the point of
breakdown of adiabaticity and the point of start of vacuum propagation
is a significant fraction of solar radius. In such a case, the calculation
of $M_{12}$ is complicated and depends on the density profile of the sun 
beyond $0.9 R_s$. Let us calculate $M_{12}$ in a simplified situation
where we will assume that the off-diagonal terms are much larger than the 
diagonal terms between the point of breakdown of adiabaticity $(t_1)$ and 
the point of restoration of adiabaticity $(t_2)$. That is, in the region 
where adiabatic approximation is not valid, the evolution is extremely  
non-adiabatic. Then the evolution equation is 
\begin{equation}
i \frac{d}{dt} \left[ \begin{array}{c} \nu_{1m} \\ \nu_{2m} 
\end{array} \right]
= \frac{1}{4E} \left[ \begin{array}{cc} 
 0 & -4 i E \dot{\theta}_m (t) \\
4 i E \dot{\theta}_m (t) & 0 \end{array} \right]
\left[ \begin{array}{c} \nu_{1m} \\ \nu_{2m} 
\end{array} \right].
\label{eq:naeqprop}
\end{equation}
Integrating this equation from $t_1$ to $t_2$, we get 
\begin{equation}
M_{12} = \sin \left( \theta (t_1) - \theta (t_2) \right). \label{eq:m12}
\end{equation}
As mentioned above, until the time $t_1$, $A >> \delta$, hence
$\theta (t_1) = \pi/2$. At time $t_2$, vacuum propagation starts,
hence $\theta (t_2) = \theta$. So we get $M_{12} = \cos \theta$,
which reduces $P_{ee}$ to vacuum survival probability. Note that
we have not made any assumption about how the density varies to
zero. The width of non-adiabatic region (given by $(t_1 - t_2)$) 
is irrelevant. In case $\theta \leq \pi/4$ there may be a point
within this non-adiabatic region where the resonance
condition is satisfied. However, the discussion above is completely 
independent of whether a resonance exists or not. The reason for this
is that the breakdown of adiabaticity condition is independent of the
existence of resonance. This is in contrast to the usual MSW effect
where the adiabaticity condition breaks down only in the neighbourhood
of a resonance.  

The above simplication is not very good because, just 
after the breakdown of adiabaticity and also just before the 
restoration of the adiabaticity, the diagonal and off-diagonal terms
in the evolution matrix are comparable. We need to consider these
regions of `moderate non-adiabaticty' separately. 
To study how good the above assumption of extreme non-adiabaticity
throughout the region of non-adiabaticity is, we define a quantity
Non-Adiabaticity Paramater (NAP). It is the ratio of the non-diagonal
term to the diagonal term in the evolution matrix (Eq.~(\ref{eq:eqprop})
\begin{eqnarray}
NAP & = & \frac{ 4 E \dot{\theta}_m}{\delta_m} \nonumber \\
& = & \frac{2 E \delta \sin 2 \theta \dot{A}}{\delta^3_m}.
\end{eqnarray}
Note that NAP is proportional to $d \rho/dt$ and is much smaller than $1$
for regions of adiabaticity and is comparable to or greater than $1$ in the 
region where adiabatic approximation breaks down. The region where NAP is 
much greater than $1$ is the region of extereme non-adiabaticity. Suppose 
density profile varies smoothly upto the edge of the sun and then abruptly
falls to zero. If the adiabatic approximation holds till the edge of the 
sun, the graph of NAP {\it vs} r looks like a Dirac delta function
at $r = R_s$. In the discussion below, we consider two different 
density profiles.  
\begin{itemize}
\item
Exponential density fall (given in Eq.~(\ref{eq:expden})) for all $r$,
even beyond $r = 0.9 R_s$.
\item
Exponential density fall upto $r = 0.9 R_s$ and then a linear density 
fall beyond $r = 0.9 R_s$. The equation of the linear fall is determined
by requiring that the value of the density and its first derivative
should match at $r = 0.9 R_s$. 
\end{itemize}
We see that for linear fall, NAP is zero until $r \simeq R_s$ and then
rises sharply at $r = R_s$,
which resembles the NAP graph for an abrupt density change.
Hence for linear fall at the edge of the sun, 
the just-so oscillation probability 
reduces to the vacuum oscillation probability. The region of moderate 
non-adiabaticity is very narrow ($r = 0.957 R_s$ to $r = 0.991 R_s$).

However for exponential fall, NAP rises slowly, becomes greater than 
$0.3$ (breakdown of adiabatic approximation) at $r = 0.96 R_s$, 
crosses $10$ around
$r = 1.12 R_s$  and falls below $10$ again for $r = 1.71 R_s$. The adiabatic
approximation (NAP < $0.3$) is restored only for $r = 2.04 R_s$.
We see that regions of moderate non-adiabaticity are reasonably large
both during increasing NAP $(0.16 R_s)$ and during decreasing NAP $0.33 R_s$. 
The transition amplitude between the two edges of 
extreme non-adiabaticity can be easily calculated, as shown in
Eqs.~(\ref{eq:naeqprop}) and~(\ref{eq:m12}).  
However, the effects of propagation through the regions of moderate
non-adiabaticity $(0.3 < {\rm NAP} < 10)$ must be taken into account.
Since the regions of moderate non-adiabaticity are wide, these can
be calculataed only by numerical integrattion of the evolution 
equation in these regions. These moderate non-adiabatic effects can
lead to the energy dependence of the just-so survival probability to be  
different from that of the vacuum survival probability. 
This investigation is in progress.

As mentioned above, for exponential density fall for all $r$, 
the survival probability 
for just-so oscillations is likely to be different from that of vacuum 
oscillations. Recently a calculation of such a modified survival 
probability was presented \cite{berkeley}. 
We find the calculation in \cite{berkeley} untenable 
for the following reasons.
\begin{enumerate}
\item
The calculation uses an expression for non-adiabatic jump
probability which was calculated under the assumption that 
the physical region in the sun
of extreme non-adiabaticity is very small so that it can be taken to
be a point (namely the resonance point) and the non-adiabatic jump
is calculated only at this point. However, as can be seen from Figure~1, 
the physical region of non-adiabaticity is of the order
of solar radius, hence the assumption of non-adiabatic jump occuring
at a point is unreasonable.
\item
In ref \cite{berkeley} it was pointed out that the survival probability 
is asymmetric under the exchange $\theta \leftrightarrow \pi/2 - \theta$.
The expression for survival probability uses a standard formula for
non-adiabatic jump probability. However, this formula is derived 
assuming the existance of a resonance \cite{petcov}.
Hence it cannot be used for scenarios where $\theta > \pi/4$.
\end{enumerate}

\section{Conclusions}
\begin{itemize}
\item
If the solar density abruptly falls to zero at the edge of the sun,
just-so oscillation probability reduces to vacuum oscillation probability.
\item
If the density fall is smooth, the just-so oscillation probability 
reduces to vacuum oscillation probability if one assumes that the
evolution is extremely non-adiabatic throughout the region where adiabatic
condition is not valid. This conclusion is independent of the exact 
density profile near the edge of the sun or the exact width of the
region of non-adiabaticity.
\item
The regions of moderate non-adiabaticity can give corrections to the
simple vacuum oscillation probability. However, for linear density 
fall these corrections are negligible.
\item
If the recoil spectrum analysis of Super Kamiokande points towards
a departure from the simple vacuum oscillation probability for just-so 
parameters, these departures can be used to obtain information on
the density profile at the edge of the sun.
\end{itemize}

Acknowledgements: We thank Prof. Paul Langacker for a 
helpful discussion.

\begin{figure}
{\centering \includegraphics{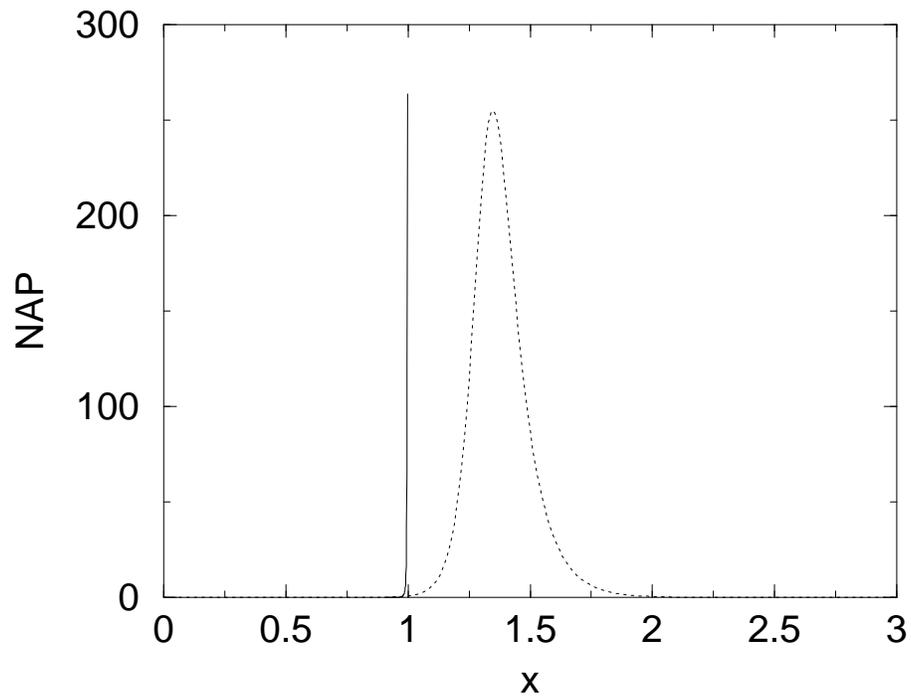} \par}
\caption{ Plot of NAP {\it vs} x $(=r/R_s)$, for linear density fall
(solid line) and exponential density fall (dotted line).}
\end{figure}

\end{document}